\documentclass[12pt]{article}
\usepackage{mathpazo}
\usepackage[latin9]{inputenc}
\usepackage[a4paper]{geometry}
\usepackage{xcolor}
\usepackage{units}
\usepackage{amsmath}
\usepackage{amssymb}
\usepackage{esint}
\PassOptionsToPackage{normalem}{ulem}
\usepackage{ulem}

\makeatletter

\providecolor{lyxadded}{rgb}{0,0,1}
\providecolor{lyxdeleted}{rgb}{1,0,0}

\DeclareRobustCommand{\lyxsout}[1]{\ifx\\#1\else\sout{#1}\fi}

\usepackage{color}
\usepackage{tikz}
\usepackage{tikz-3dplot}
\usetikzlibrary{patterns}
\usepackage{ifthen}
\usepackage{pgfplots}
\usepackage{palatino}

\usepackage{calrsfs}
\usepackage[outerbars]{changebar}
\newcommand{\abs}[1]{\lvert#1\rvert}
\newcommand{\avg}[1]{\langle #1 \rangle}

\usepackage{units}
\usepackage{authblk}
\makeatother

\begin{document}

\title{Geometry of Uncertainty Relations for\\
 Linear Combinations of Position and Momentum}
\author[1]{Spiros Kechrimparis\thanks{skechrimparis@gmail.com}}
\author[2]{Stefan Weigert\thanks{stefan.weigert.@york.ac.uk}}
\affil[1]{Department of Applied Mathematics, Hanyang University
	(ERICA), \protect\\  55 Hanyangdaehak-ro, Ansan, Gyeonggi-do, 426-791, Korea}
\affil[2]{Department of Mathematics, University of York \protect\\
	York, YO10 5DD, United Kingdom }

\maketitle

\begin{abstract}
For a quantum particle with a single degree of freedom, we derive
preparational sum and product uncertainty relations satisfied by $N$
linear combinations of position and momentum observables. The state-independent bounds
depend on their \emph{degree of incompatibility} defined by the area
of a parallelogram in an $N$-dimensional coefficient space. Maximal
incompatibility occurs if the observables give rise to regular polygons
in phase space. We also conjecture a Hirschman-type uncertainty relation
for $N$ observables linear in position and momentum, generalizing
the original relation which lower-bounds the sum of the position and
momentum Shannon entropies of the particle.

\global\long\def\bk#1#2{\langle#1|#2\rangle}

\global\long\def\kb#1#2{|#1\rangle\,\langle#2|}

\global\long\def\braket#1#2{\langle#1|#2\rangle}

\global\long\def\ket#1{|#1\rangle}

\global\long\def\bra#1{\langle#1|}

\global\long\def\c#1{\mathbb{C}^{#1}}

\global\long\def\abs#1{\mid#1\mid}

\global\long\def\avg#1{\langle#1\rangle}
 
\end{abstract}

\section{Introduction}

For a long time, quantum mechanical uncertainty relations were tantamount
to statements about \emph{pairs }of non-commuting observables. Heisenberg's
discussion of a fictitious $\gamma$-ray microscope in 1927 \cite{heisenberg27}
led Kennard to immediately derive a rigorous \emph{preparational }uncertainty
relation \cite{kennard27} for the product of the variances of position
and momentum observables. The existence of pairwise incompatible observables
represents one of the defining features of quantum theory. 

It is natural to suspect that similar limitations may also exist for
triples, quadruples, etc. of non-commuting observables, and they may
not be reducible to uncertainty relations for pairs. Indeed, the triple
uncertainty relation \cite{kechrimparis+14}, for example, 
\begin{equation}
\Delta p\Delta q\Delta r\geq\left(\tau\frac{\hbar}{2}\right)^{\nicefrac{3}{2}}\,,\quad\tau=\csc\left(\frac{2\pi}{3}\right)\simeq1.15\,,\label{eq: triple UR}
\end{equation}
bounds the product of the variances of \emph{three} \emph{pairwise
canonical} operators, $\hat{p},\hat{q}$, and $\hat{r}=-\hat{p}-\hat{q}$.
The bound \eqref{eq: triple UR} follows neither from individually
applying Heisenberg's uncertainty relation to each of the canonical
pairs of observables $(\hat{p},\hat{q})$, $(\hat{q},\hat{r})$ and
$(\hat{r},\hat{p})$, nor from its generalization found by Robertson
and Schrödinger \cite{robertson29,schroedinger30}. Early on, Robertson
derived inequalities for sets of more than two observables \cite{robertson34}
but the results do not cover the situation we will consider. For example,
his bound on the product of the variances of the observables $\hat{p},\hat{q}$,
and $\hat{r}$ turns out to be the trivial one, $\Delta p\Delta q\Delta r\geq0$.

For a long time, uncertainty relations for continuous variables were
thought to be of mainly conceptual interest. For systems with more
than one degree of freedom, however, they are now known to provide
tools to detect entanglement. The criteria may, for example, use the variances of position and momentum operators only, as in \cite{duan+00},
or the entire covariance matrix \cite{guhne+2003}. Not surprisingly,
the triple uncertainty relation \eqref{eq: triple UR} also lends
itself to detect entanglement, according to a recent proposal and
its quantum optical realization \cite{paul+16}.

In this paper, we will derive tight inequalities for the product and
the sum of variances of finitely many observables for a single continuous
variable describing, for example, a quantum particle restricted to
move on the real line. We limit ourselves to \emph{linear combinations}
of position and momentum observables. Recent work on uncertainty relations
beyond pairs of observables \cite{chen+15,qin+16} has led to new
\emph{state-dependent} bounds, as well as to bounds on the variances
of multiple \emph{unitary} operators \cite{bagchi+16}. In contrast
to these approaches, the \emph{linearity} of the observables we consider
will lead to \emph{state-independent} bounds, for the traditional
case of Hermitean observables.

We will also introduce a many-observable generalization of the \emph{entropic}
uncertainty relation conjectured by Hirschman \cite{hirschman57}
in 1957 (but proved only two decades later \cite{bialnicki+75,beckner75}).
It is, in fact, straightforward to ask for a bound on the sum of \emph{more
than two} Shannon entropies for a given quantum state. As for variance-based
uncertainty relations, we again expect Gaussian states to play an
important role, suggested by the fact that they \emph{saturate} the
proposed inequalities. Recent results for entropic uncertainty relations
valid in Hilbert spaces of small finite dimensions show how difficult
it is to obtain tight bounds \cite{riccardi+17}.

We have laid out this paper in the following way. In Sec. \ref{Finite}
we derive inequalities obeyed by the variances of $N$ observables
linear in position and momentum. The ``non-commutativity'' encoded
in their pairwise commutators can be expressed in the \emph{degree
of incompatibility}, i.e. a real number which determines the lower
bounds on sums and products of variances. Geometrically, this degree
is given by the area of a suitably defined parallelogram in coefficient
space $\mathbb{R}^{N}$. Specific sets of observables associated with
regular polygons are shown to saturate the bounds. In Sec. \ref{Entropies}
we generalize Hirschman's entropic uncertainty relation to more than
two observables and explain why we expect the conjectured bounds to
be tight. The last section summarizes our results and we will draw
conclusions.

\section{Variance-based uncertainty relations \label{Finite}}

The position and momentum observables $\hat{p}$ and $\hat{q}$ of
a single quantum particle act on a Hilbert space the elements of which
can be represented as square-integrable functions over the real line,
i.e. ${\cal H}=L^{2}(\mathbb{R})$. We introduce $N$ Hermitean operators
by combining them linearly, 
\begin{equation}
\hat{r}_{j}=a_{j}\hat{p}+b_{j}\hat{q}\,,\qquad a_{j},b_{j}\in\mathbb{R}\,,\qquad j=1\ldots N\,.\label{eq: operator rj}
\end{equation}
To exclude a trivial situation, at least two of the operators $\hat{r}_{j},j=1\ldots N$,
should not commute. Using a system of units in which both position
and momentum have physical dimension $\sqrt{\hbar}$, the coefficients
$a_{j},b_{j}$, $j=1\ldots N$, are dimensionless. The operators in
\eqref{eq: operator rj} represent \emph{observables} since they can
be measured as quadratures of an electromagnetic field in quantum-optical
experiments, for example \cite{welsch+09,manko+09,bellini+2012}.
Each observable $\hat{r}_{j}$ is characterized by a vector in a two-dimensional
Euclidean space, 
\begin{equation}
\mathbf{r}_{j}=\left(\begin{array}{c}
a_{j}\\
b_{j}
\end{array}\right)\in\mathbb{R}^{2}\,,\qquad j=1\ldots N\,.\label{eq: define rj vectors}
\end{equation}
We will call $r_{j}=\sqrt{a_{j}^{2}+b_{j}^{2}}$ the ``length''
of the observable $\hat{r}_{j}$.

The fundamental commutation relation 
\begin{equation}
\left[\hat{p},\hat{q}\right]=\frac{\hbar}{i}\hat{I}\,, \label{eq: ipq=00003D00003Dhbar}
\end{equation}
where $\hat{I}$ is the the identity operator on the Hilbert space
${\cal H}$, implies that the pairwise commutators of the \textbf{$r$}-observables
are given by 
\begin{equation}
\left[\hat{r}_{j},\hat{r}_{k}\right]=A(\mathbf{r}_{j},\mathbf{r}_{k})\,\frac{\hbar}{i}\hat{I}\,,\qquad j,k=1\ldots N\,,\label{eq: rj commutator}
\end{equation}
with the function $A(\cdot,\cdot)$ computing the (signed) area of
the parallelogram determined by the vectors $\mathbf{r}_{j},\mathbf{r}_{k}\in\mathbb{R}^{2}$,
\begin{equation}
A(\mathbf{r}_{j},\mathbf{r}_{k})=\left(a_{j}b_{k}-a_{k}b_{j}\right)\equiv A_{jk}\,.\label{eq:jk-area}
\end{equation}
Thus, the commutation relations between the $r$-observables are encoded
in the $N$-by-$N$ skew-symmetric matrix $\mathbf{A}$ with matrix
elements $A_{jk}=-A_{kj}$. This antisymmetric structure finds its
natural expression in a coordinate-in\-de\-pen\-dent formulation.
Let us treat the linear combinations $\hat{r}_{j}$, $j=1\ldots N$,
as components of a vector operator with $N$ components, 
\begin{equation}
\hat{\mathbf{r}}=\left(\begin{array}{c}
\hat{r}_{1}\\
\vdots\\
\hat{r}_{N}
\end{array}\right)\equiv\mathbf{a}\hat{p}+\mathbf{b}\hat{q}\,,\qquad\mathbf{a},\mathbf{b}\in\mathbb{R}^{N}\,.\label{eq: r via a and b}
\end{equation}
Since the components of the exterior product of two vectors $\mathbf{u},\mathbf{v}\in\mathbb{R}^{N}$,
are given by 
\begin{equation}
\left(\mathbf{u}\wedge\mathbf{v}\right)_{jk}=u_{j}v_{k}-u_{k}v_{j}\,,\quad j,k=1\ldots N\,,\label{eq: def wedge}
\end{equation}
we find that the $N^{2}$ commutation relations \eqref{eq: rj commutator}
elegantly combine to 
\begin{equation}
\hat{\mathbf{r}}\wedge\hat{\mathbf{r}}=\,\mathbf{a}\wedge\mathbf{b}\,\frac{\hbar}{i}\hat{I}\,.\label{eq: rr =00003Da wedge b}
\end{equation}
Normally, the wedge product of a vector with itself is equal to zero
but this does not apply to the left-hand-side of \eqref{eq: rr =00003Da wedge b}
because $\hat{\mathbf{r}}$ is a vector with operator-valued, \emph{non-commuting
}components. The relation is consistent with writing $\hat{\mathbf{r}}=\sum_{j=1}^{N}\hat{r}_{j}\mathbf{e}_{j}$,
where the vectors $\mathbf{e}_{j}$, $j=1\ldots N$, form the standard
orthonormal basis of the space $\mathbb{R}^{N}$, and using the anti-symmetry
of the exterior products $\mathbf{e}_{j}\wedge\mathbf{e}_{k}=-\mathbf{e}_{k}\wedge\mathbf{e}_{j}$.
In a similar spirit, the commutation relations for a spin, $\sum_{qr}\varepsilon_{pqr}\hat{s}_{q}\hat{s}_{r}=i\hat{s}_{p}$,
$p,q,r\in\left\{ x,y,z\right\} $, can be written formally as a cross
product, $\hat{\boldsymbol{s}}\times\hat{\boldsymbol{s}}=i\hbar\hat{\boldsymbol{s}}$,
by combining the three operator-valued components of a quantum spin
in a single vector $\hat{\boldsymbol{s}}=\frac{\hbar}{2}\hat{\boldsymbol{\sigma}}$
(see \cite{baym69}, for example). It will be useful to write Eq.
\eqref{eq: def wedge} in vector form, i.e. $\mathbf{u}\wedge\mathbf{v}=\mathbf{u}\otimes\mathbf{v}-\mathbf{v}\otimes\mathbf{u}$,
where the outer product $\mathbf{u\otimes\mathbf{v}}$ of two vectors
is defined by 
\begin{equation}
\left(\mathbf{u\otimes\mathbf{v}}\right)_{jk}\equiv\left(\mathbf{u\mathbf{v}}^{T}\right)_{jk}\,,\qquad j,k=1\ldots N\,.\label{eq: def outer product}
\end{equation}

The squared norm or \emph{magnitude }of the bi-vector $\mathbf{a}\wedge\mathbf{b}\in\bigwedge^{2}(\mathbb{R})$
is given by

\begin{equation}
\left|\mathbf{a}\wedge\mathbf{b}\right|^{2}=\sum_{j>k=1}^{N}A_{jk}^{2}\,.\label{eq: squared norm for bivector}
\end{equation}
It has a simple expression in terms of the vectors defining the $r$-operators,
\begin{equation}
\left|\mathbf{a}\wedge\mathbf{b}\right|^{2}=\sum_{j>k=1}^{N}\left(a_{j}b_{k}-a_{k}b_{j}\right)^{2}=\left|\mathbf{a}\right|{}^{2}\,\left|\mathbf{b}\right|{}^{2}-\left(\mathbf{a}\cdot\mathbf{b}\right)^{2}\,,\label{eq: norm via a and b}
\end{equation}
which follows from Lagrange's identity for real numbers. Using $\mathbf{a}\cdot\mathbf{b}=|\mathbf{a}|\:|\mathbf{b}|\cos\phi$,
where $\phi\in[0,\pi)$ is the angle between the vectors $\mathbf{a}$
and $\mathbf{b}$, one finds 
\begin{equation}
\left|\mathbf{a}\wedge\mathbf{b}\right|=|\mathbf{a}|\:|\mathbf{b}|\sin\phi\,,\label{eq: length of wedge product}
\end{equation}
in agreement with the wedge product being a generalization of the
vector product in $\mathbb{R}^{3}$.

Geometrically, the squared norm of a bi-vector $\mathbf{a}\wedge\mathbf{b}$
is given by the sum of the squared areas of the parallelograms defined
by all pairs of vectors $\mathbf{r}_{j}\in\mathbb{R}^{2}$, $j=1\ldots N$,
which, according to \eqref{eq: norm via a and b}, equals the square
of the area of the parallelogram spanned by the vectors $\mathbf{a},\mathbf{b}\in\mathbb{R}^{N}$
in coefficient space. Not surprisingly, the norm is also closely related
to a norm of the antisymmetric matrix $\mathbf{A}$ defined by Eq.
\eqref{eq: rj commutator}: the square of its Frobenius (or Hilbert-Schmidt
or $L_{2,2})$ norm reads 
\begin{equation}
\left\Vert \mathbf{A}\right\Vert {}_{F}^{2}=\text{\mbox{Tr}}\left(\mathbf{A}^{T}\mathbf{A}\right)=\sum_{j,k=1}^{N}A_{jk}^{2}=2\sum_{j>k=1}^{N}A_{jk}^{2}=2\left|\mathbf{a}\wedge\mathbf{b}\right|^{2}\,.\label{eq: Frobenius and bivector norm}
\end{equation}
This relation will be used in Sec. \ref{subsec:Degrees-of-incompatibility}.

\subsection{Sum and product inequalities \label{NObservables}}

The variances $\Delta^{2}r_{j}\equiv\bra{\psi}\hat{r}_{j}^{2}\ket{\psi}-\bra{\psi}\hat{r}_{j}\ket{\psi}^{2}$
of the $N$ linearly dependent $r$-observables in a pure state $\ket{\psi}\in{\cal H}$
are given by 
\begin{equation}
\Delta^{2}r_{j}=a_{j}^{2}\Delta^{2}p+b_{j}^{2}\Delta^{2}q+2a_{j}b_{j}C_{pq}\,,\qquad j=1\ldots N\,,\label{eq: variance of rj}
\end{equation}
where we have introduced the covariance

\begin{equation}
C_{pq}=\frac{1}{2}\left(\bra{\psi}\left(\hat{p}\hat{q}+\hat{q}\hat{p}\right)\ket{\psi}-\bra{\psi}\hat{p}\ket{\psi}\bra{\psi}\hat{q}\ket{\psi}\right)\,.\label{eq: correlation term}
\end{equation}
Adding the variances $\Delta^{2}r_{j}$, we obtain 
\begin{align}
\sum_{j=1}^{N}\Delta^{2}r_{j} & =|\mathbf{a}|^{2}\Delta^{2}p+|\mathbf{b}|^{2}\Delta^{2}q+2\,\mathbf{a}\cdot\mathbf{b}\,C_{pq}\,.\label{eq: sum of variances}
\end{align}
The right-hand-side of Eq. \eqref{eq: sum of variances} defines a
functional of three operators quadratic in position and momentum.
The bounds of such expressions have been studied systematically in
\cite{kechrimparis+16}. An explicit, non-trivial lower bound has
been obtained for the linear combination of the variances $\Delta^{2}p,\Delta^{2}q$
and the covariance $C_{pq}$ (see Eq. (72) of \cite{kechrimparis+16}),
\begin{equation}
\mu\Delta^{\!2}p+\nu\Delta^{\!2}q+2\lambda C_{pq}\geq\hbar\sqrt{\mu\nu-\lambda^{2}},\qquad\mu,\nu>0\,,\quad\mu\nu>\lambda^{2}\,.\label{eq:linear UR}
\end{equation}
This inequality follows directly and elegantly from the non-negative
expectation value of a quadratic form in position and momentum in
an arbitrary quantum state $\hat{\rho}$, 
\begin{equation}
\text{Tr\ensuremath{\left[\hat{z}\,\hat{\rho}\,\hat{z}^{\dagger}\right]}=\text{Tr \ensuremath{\left[\left(\hat{z}\,\hat{\rho}^{\nicefrac{1}{2}}\right)\left(\hat{z}\,\hat{\rho}^{\nicefrac{1}{2}}\right)^{\dagger}\right]}}\ensuremath{\geq}0\,,}\label{eq: positive trace expression}
\end{equation}
where 
\begin{equation}
\hat{z}=\alpha\left(\hat{p}-\langle\hat{p}\rangle\right)+\beta\left(\hat{q}-\langle\hat{q}\rangle\right)\,,\quad\langle\hat{p}\rangle\equiv\text{Tr}\left[\hat{\rho}\,\hat{p}\right]\,,\quad\text{etc.}\,,\label{eq: definition of z}
\end{equation}
and $\alpha,\beta\in\mathbb{R}$, are complex numbers which satisfy
$\text{Im}\left(\alpha\,\beta^{*}\right)\geq0$. A straightforward
calculation shows that upon identifying $\mu\equiv\left|\alpha\right|^{2},\nu=\left|\beta\right|^{2}$
and $\lambda=\text{Re}\left(\alpha^{*}\,\beta\right)$ one obtains
indeed \eqref{eq:linear UR}, valid for both pure \emph{and }mixed
states.

Setting 
\begin{equation}
\mu\equiv|\mathbf{a}|^{2}\,,\qquad\nu\equiv|\mathbf{b}|^{2}\,,\qquad\lambda\equiv\mathbf{a}\cdot\mathbf{b}\,,\label{eq: parameter values}
\end{equation}
we can apply the tight inequality \eqref{eq:linear UR} since $|\mathbf{a}|^{2},|\mathbf{b}|^{2}>0$
and $|\mathbf{a}|^{2}\,|\mathbf{b}|^{2}>\left(\mathbf{a}\cdot\mathbf{b}\right)^{2}$
hold. Recalling the identity \eqref{eq: norm via a and b} leads to
the \emph{sum inequality} for arbitrary quantum states, 
\begin{equation}
\sum_{j=1}^{N}\Delta^{2}r_{j}\geq\hbar\left|\mathbf{a}\wedge\mathbf{b}\right|\,,\label{eq: SumNG}
\end{equation}
which is our first main result. Appendix A presents an alternative
derivation which is based on the validity of \eqref{eq:linear UR}
for pure states and the concavity of the variance.

Eq. \eqref{eq: SumNG} correctly reproduces both the pair and triple
sum identities leading to the bounds $\hbar$ and $\hbar\sqrt{3}$,
respectively. The bound is \emph{state-independent} because the commutator
between any two linear combinations of position and momentum is a
scalar multiple of the identity. A trivial bound (zero) is obtained
if the inequality \eqref{eq:linear UR} is applied to each term of
the sum \eqref{eq: sum of variances} separately, i.e. \emph{before}
instead of \emph{after} the summation in \eqref{eq: sum of variances}.

Using \eqref{eq: rj commutator} it is possible to express the lower
bound of the inequality \eqref{eq: SumNG} in terms of the pairwise
commutators between the $N$ operators,
\begin{equation}
\left(\sum_{j=1}^{N}\Delta^{2}r_{j}\right)^{2}\geq\sum_{j>k=1}^{N}\left|\avg{[\hat{r}_{j},\hat{r}_{k}]}\right|^{2}\,,\label{eq: bound as commutators}
\end{equation}
where the expectation values of the commutators are taken in the state
$\ket{\psi}$. Thus, the sum of the variances of $N$ different linear
combinations $\hat{r}_{j}$ of position and momentum operators is
seen to be bounded from below by the square root of the \emph{sum
of the modulus squared of all commutators between the operators}.
Applying the Cauchy-Schwarz inequality to this expression for $N>2$,
we find that 

\begin{equation}
\sum_{j>k=1}^{N}\left|\avg{[\hat{r}_{j},\hat{r}_{k}]}\right|^{2}>\left(\frac{1}{N-1}\sum_{j>k=1}^{N}\left|\avg{[\hat{r}_{j},\hat{r}_{k}]}\right|\right)^{2}\,.\label{comparison of bounds }
\end{equation}
Upon concatenating this inequality with \eqref{eq: bound as commutators},
we obtain a bound on the sum of $N$ variances which can be derived
directly from the inequalities valid for each of the $N(N-1)$ pairs
$(\Delta^{2}r_{k}+\Delta^{2}r_{j})$, $1\leq k<j\leq N$. The stronger
bound \eqref{eq: SumNG} shows that these uncertainty relations for
$N$ observables do \emph{not} follow from those of the pairwise inequalities.
According to \cite{trifonov02}, the concatenated inequality is actually
known to hold for \emph{arbitrary} observables $\hat{r}_{j}$, $j=1\ldots N$,
not just linear combinations of position and momentum. However, it
is also not tight as the case of \emph{three} observables shows \cite{kechrimparis15,song+16}.

To identify the states saturating the inequality \eqref{eq: SumNG}
let us introduce the ground state $\ket 0$ of a harmonic quantum
oscillator with unit mass and frequency, and the family of coherent
states $\ket{\alpha}=\hat{T}_{\alpha}\ket 0$, where the unitary operator
\begin{equation}
\hat{T}_{\alpha}=\exp\left[i\left(p_{0}\hat{q}-q_{0}\hat{p}\right)/\hbar\right]\,,\qquad\alpha=\frac{1}{\sqrt{2\hbar}}\left(q_{0}+ip_{0}\right)\,,\label{TranslOp}
\end{equation}
generates a position and momentum translation by amounts $q_{0}$
and $p_{0}$. As shown in \cite{kechrimparis+16}, the inequality
\eqref{eq:linear UR} and hence the sum inequality \eqref{eq: SumNG}
attain their minimum if the oscillator resides in a suitably \emph{squeezed
}ground state $\ket 0$, 
\begin{equation}
\ket{\mu,\nu,\lambda}=\hat{G}_{\frac{\lambda}{\nu}}\hat{S}_{\frac{1}{2}\ln\left(\frac{\nu}{\sqrt{\mu\nu-\lambda^{2}}}\right)}\ket 0\,,\label{eq: minima of linear ineq}
\end{equation}
or in any state obtained from rigidly displacing it, i.e. $\hat{T}_{\alpha}\ket{\mu,\nu,\lambda}$.
Here, the unitary operator 
\begin{equation}
\hat{G}_{g}=\exp\left[ig\hat{p}^{2}/2\hbar\right]\,,\qquad g\in\mathbb{R}\,,\label{eq: gauge operator}
\end{equation}
generates a \emph{momentum gauge} transformation while 
\begin{equation}
\hat{S}_{\gamma}=\exp\left[i\gamma\left(\hat{q}\hat{p}+\hat{p}\hat{q}\right)/2\hbar\right]\,,\qquad\gamma\in\mathbb{R}\,,\label{Squeeze}
\end{equation}
\emph{squeezes }a state along the coordinate axes of phase space.
For $N=2$, with observables $\hat{r}_{1}=\hat{p}$ and $\hat{r}_{2}=\hat{q}$,
say, corresponding to $\mu=\nu=1$ and $\lambda=0$, we find $\hat{G}_{0}=\hat{S}_{0}=\hat{I}$.
This result agrees with the well-known fact that the only states minimizing
the sum $\Delta^{2}p+\Delta^{2}q$ are given by the ground state of
a harmonic oscillator and its rigid displacements in phase space.

Next, we wish to generalize Heisenberg's uncertainty relation by deriving
a bound on the value of the \emph{product }of the variances for the
observables $\hat{r}_{j},j=1\ldots N$, 
\begin{equation}
J\left[\ket{\psi}\right]=\prod_{j=1}^{N}\Delta^{2}r_{j}\,,\label{eq: product functional}
\end{equation}
where $N\geq2$. Using the identities \eqref{eq: variance of rj},
the functional $J\left[\ket{\psi}\right]$, which associates a number
to each state $\ket{\psi}$, turns into a polynomial of order $N$
in the basic variances $\Delta^{2}p,\Delta^{2}q$, and the covariance
$C_{pq}$. Its lower bound could be determined by applying the method
described in \cite{kechrimparis+16}. However, in this highly symmetric
case, another method turns out to be simpler which enables us to minimize
the product $J$ while respecting the constraint given by the sum
inequality \eqref{eq: SumNG}.

A function $J(\vec{x})$ has a minimum in the presence of an inequality
$g(\vec{x})\leq0$ if the Karush-Kuhn-Tucker (KKT) conditions \cite{ruszczynski+06}
are satisfied, 
\begin{align}
\frac{\partial J(\vec{x})}{\partial x_{j}}+\kappa\frac{\partial g(\vec{x})}{\partial x_{j}}=0\,, & \qquad j=1\ldots N\,,\label{KKTcond}\\
\kappa g(\vec{x})=0\,,\label{eq: KKTcond b}
\end{align}
where $\kappa$ is a positive constant yet to be determined. Identifying
the variables $x_{j}$ with the variances $\Delta^{2}r_{j},j=1\ldots N$,
the constraint \eqref{eq: SumNG} reads $g(\vec{x})\equiv c-\sum_{j}x_{j}\leq0$,
with the positive number $c=\hbar\left|\mathbf{a}\wedge\mathbf{b}\right|$.

The unique solution of the KKT conditions \eqref{KKTcond} is easily
found to be 
\begin{equation}
x_{1}=x_{2}=\ldots=x_{N}=\frac{c}{N}\,,\label{eq: KKT solutions}
\end{equation}
which implies that the smallest value of the functional $J\left[\ket{\psi}\right]$
is given by $\left(c/N\right)^{N}$. In terms of the original variables,
we finally obtain the \emph{product inequality} for the variances
of $N$ linear combinations of position and momentum, 
\begin{equation}
\prod_{j=1}^{N}\Delta^{2}r_{j}\geq\left(\frac{\hbar\left|\mathbf{a}\wedge\mathbf{b}\right|}{N}\right)^{N}\,,\label{ProdNG}
\end{equation}
our second main result. The lower bounds for Heisenberg's uncertainty
relation and for the triple product uncertainty relation \eqref{eq: triple UR}
are reproduced correctly. The bounds are symmetric in all pairs of
the $N$ observables $\hat{r}_{j},j=1\ldots N$, and they display
a neat structure which involves the exterior product of the momentum
and position coefficients in $\mathbb{R}^{N}$. The result \eqref{ProdNG}
is genuinely different from Robertson's inequalities for $N$ observables
\cite{robertson34} since already in the case of $N=3$ only a trivial
bound results, $\Delta p\Delta q\Delta r\geq0$. The
derivation of inequality \eqref{ProdNG} also applies to mixed states,
i.e. $\Delta^{2}r_{j}=\text{Tr}\left(\hat{\rho}\hat{r}_{j}^{2}\right)-\left(\text{Tr}\left(\hat{\rho}\hat{r}_{j}\right)\right)^{2},j=1\ldots N$.

\subsection{Regular polygons \label{NCanonical}}

\begin{figure}[t]
\centering{}\tikzset{>=latex} \begin{tikzpicture}[scale=1.4]
\node at (95:2.2)  {$\mathbf{r}_1$};
\node at (104:1)  {$R_{5}$};
\node at (54:2.05)  {$\nicefrac{A_{21}}{2}$};
\node at (163:2.3) {$\mathbf{r}_5$};
\node at (234:2.3) {$\mathbf{r}_4$};
\node at (306:2.3) {$\mathbf{r}_3$};
\node at (17:2.3) {$\mathbf{r}_2$};
\node at (85:2.5) {$\mathbf{p}$};
\node at (-6:2.4) {$\mathbf{q}$};
\coordinate (r2) at (1.9,0.61);
\coordinate (r1) at (0,2);
\coordinate (or) at (0,0); 
\draw [ ->] [black] plot [smooth, tension=1] coordinates { (53:1.9) (52:1.68) (48:1.73) (48:1.4)};
\draw[pattern=north east lines] (r2) -- (r1) -- (or);
\draw[dashed] (0,2) -- (1.9,0.61);
\draw[dashed] (0,2) -- (-1.9,0.61);
\draw[dashed] (-1.9,0.61) -- (-1.17,-1.61);
\draw[dashed] (1.9,0.61) -- (1.17,-1.61);
\draw[dashed] (-1.17,-1.61) -- (1.17,-1.61);
\draw [ ->] [line width=0.5mm] (0,0) -- (90:2);
\draw [ ->] [line width=0.5mm] (0,0) -- (162:2);
\draw [ ->] [line width=0.5mm] (0,0) -- (234:2);
\draw [ ->] [line width=0.5mm] (0,0) -- (306:2);
\draw [ ->] [line width=0.5mm] (0,0) -- (18:2);
\draw [ ->] (0,-2.3) -- (0,2.7);
\draw [ ->] (-2.5,0) -- (2.5,0);	\end{tikzpicture} \caption{A regular pentagon in the dimensionless ``phase space'' $\mathbb{R}^{2},$
associated with the canonical operators $\hat{r}_{j},j=1\ldots5$,
introduced in \eqref{eq: rjs for polygons}, of circumradius $R_{5}=1/\sqrt{\sin(2\pi/5)}$
(cf. Eq. \eqref{eq:Circumradius}) and with area $A=5/2$. The shaded
triangle has half the size of the area $A_{21}\equiv A(\mathbf{r}_{2},\mathbf{r}_{1})\equiv1$
given by the parallelogram spanned by the vectors $\mathbf{r}_{2}$
and $\mathbf{r}_{1}$ (cf. Eq. \eqref{eq:jk-area}).}
\label{Pentagon} 
\end{figure}
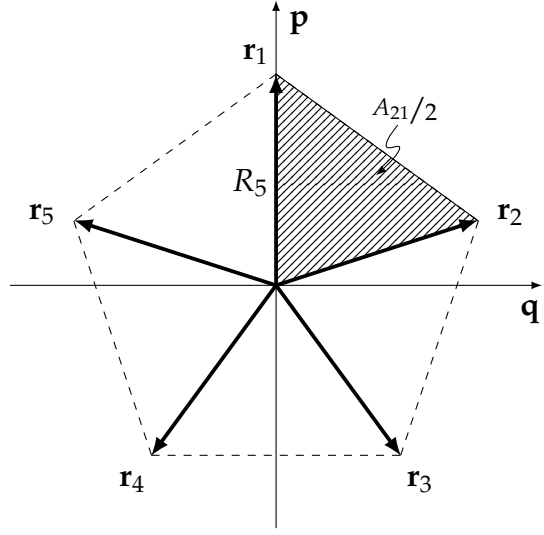

Let us now determine the bounds for $N$ observables arranged in a
symmetric way. We assume that the tips of the vectors $\mathbf{r}_{j}\in\mathbb{R}^{2}$,$j=1\ldots N$,
are located on a circle of radius $R\in(0,\infty)$, and that they
are distributed homogeneously. Explicitly, we have 
\begin{equation}
\hat{r}_{j}=\left(R\cos\varphi_{j}\right)\hat{p}+\left(R\sin\varphi_{j}\right)\hat{q}\,,\qquad\varphi_{j}=\frac{2\pi(j-1)}{N}\,,\qquad j=1,\dots,N\,.\label{eq: rjs for polygons}
\end{equation}
The tips of the vectors define a regular polygon with $N$ vertices
in the space $\mathbb{R}^{2}$, as illustrated in Fig. \eqref{Pentagon}.\emph{
}We align the first observable with the momentum operator,
i.e. $\hat{r}_{1}=R\hat{p}$. This choice is not a restriction since
the commutation relations do not change under rotations in $\mathbb{R}^{2}$
(cf. Appendix B).

From a \emph{structural} point of view, the value of the constant
$R$ is not important as it only rescales all observables. One natural
choice to fix this scale is to require that any two adjacent observables
form a canonical pair, 
\begin{equation}
[\hat{r}_{j},\hat{r}_{j+1}]=\frac{\hbar}{i}\hat{I}\,,\qquad\hat{r}_{N+1}\equiv\hat{r}_{1}\,,\qquad j=1\dots N\,.\label{eq: adjacent commutator}
\end{equation}
These conditions are satisfied if the circumradius $R$ of the polygon
takes the value 
\begin{equation}
R_{N}=\frac{1}{\sqrt{\sin\Delta_{N}}}\,,\qquad\Delta_{N}=\frac{2\pi}{N}\,.\label{eq:Circumradius}
\end{equation}
In this case, the parallelograms defined by any two consecutive vectors
$\mathbf{r}_{j}$ and $\mathbf{r}_{j+1}$, which enclose the angle
$2\pi/N$, have unit area area, $A(\mathbf{r}_{j},\mathbf{r}_{j+1})=1$.
As the angles between neighbouring vectors decrease with larger values
of $N$, the circumradius of the polygon must increase as $R_{N}\simeq\sqrt{N}$
in order to ensure \eqref{eq: adjacent commutator}.

Since the coefficient vectors $\mathbf{a}$ and $\mathbf{b}$ have
components 
\begin{equation}
a_{j}=R_{N}\cos\varphi_{j}\,,\qquad b_{j}=R_{N}\sin\varphi_{j}\,,\qquad j=1\ldots N\,,\label{eq: ab-coefficients}
\end{equation}
we find that 
\begin{equation}
\left|\mathbf{a}\wedge\mathbf{b}\right|=\frac{NR_{N}^{2}}{2}\equiv\frac{N}{2\sin\Delta_{N}}\,.\label{eq: special value of wedge prod}
\end{equation}
Here we have used the trigonometric identities 
\begin{align}
\sum_{j=1}^{N}\sin^{2}\left(\frac{2\pi j}{N}\right)=\frac{N}{2}\qquad\mbox{and}\qquad\sum_{j=1}^{N}\sin\left(\frac{4\pi j}{N}\right)=0\label{eq: trig identities}
\end{align}
to show that 
\begin{equation}
|\mathbf{a}|^{2}=|\mathbf{b}|^{2}=\frac{NR_{N}^{2}}{2}\qquad\mbox{and}\qquad\mathbf{a}\cdot\mathbf{b}=0\,,\label{eq: polygon vectors a and b}
\end{equation}
respectively. Now the identity \eqref{eq: norm via a and b} implies
that the sum and the product inequalities (see Eqs. \eqref{eq: SumNG}
and \eqref{ProdNG}) for the variances of $N$ observables associated
with regular polygons are given by 
\begin{equation}
\sum_{j=1}^{N}\Delta^{2}r_{j}\geq\frac{N\hbar}{2\sin\Delta_{N}}\qquad\mbox{and}\qquad\prod_{j=1}^{N}\Delta^{2}r_{j}\geq\left(\frac{\hbar}{2\sin\Delta_{N}}\right)^{N}\,,\label{eq: SumN}
\end{equation}
respectively .

It is possible to absorb the factor $\sin\Delta_{N}$ on the right-hand-side
of these inequalities by considering vectors $\mathbf{r}_{j}$ in
\eqref{eq: rjs for polygons} with tips located on the \emph{unit}
circle. In this case, the right-hand-side of the commutators \eqref{eq: adjacent commutator}
is found to be proportional to $\sin\Delta_{N}\simeq1/\sqrt{N}$ since
adjacent observables differ less and less for increasing values of
$N$. Then, the bounds in Eqs.\emph{ }\eqref{eq: SumN} take particularly
simple forms, 
\begin{equation}
\sum_{j=1}^{N}\Delta^{2}r_{j}\geq N\,\frac{\hbar}{2}\qquad\mbox{and}\qquad\prod_{j=1}^{N}\Delta^{2}r_{j}\geq\left(\frac{\hbar}{2}\right)^{N}\,,\label{eq: simple SumN and Prod N}
\end{equation}
i.e. each variance formally contributes at least an amount $\hbar/2$.
The states that saturate these inequalities are the \emph{coherent}
states $\ket{\alpha}=\hat{T}_{\alpha}\ket 0$, introduced via Eq.
\eqref{TranslOp}. If $N=2$ or $N=4$, the left-hand-side of the
product inequality depends only on $\Delta p\Delta q$ which is invariant
under squeezing transformations, hence leading to a \emph{larger}
family of extremal states, namely suitably squeezed states. Products
of three (or more than four) variances do \emph{not }exhibit this
continuous symmetry.

\subsection{Degrees of incompatibility\label{subsec:Degrees-of-incompatibility}}

In this section, we will argue that the dependence of the sum and
product bounds on only the norm $\left|\mathbf{a}\wedge\mathbf{b}\right|$
is not a coincidence. We will show that there exists a transformation
which maps the vector operator $\hat{\mathbf{r}}=\mathbf{a}\hat{q}+\mathbf{b}\hat{p}$
to $\hat{\mathbf{r}}^{\prime}=\mathbf{a}^{\prime}\hat{q}+\mathbf{b}^{\prime}\hat{p}$
in such a way that the commutation relations \eqref{eq: rr =00003Da wedge b}
assume their\emph{ standard form}, 
\begin{equation}
\hat{\mathbf{r}}^{\prime}\wedge\hat{\mathbf{r}}^{\prime}=\left|\mathbf{a}\wedge\mathbf{b}\right|\,\mathbf{e}_{1}\wedge\mathbf{e}_{2}\,\frac{\hbar}{i}\hat{I}\,,\label{eq: CCR in standard form}
\end{equation}
where $\mathbf{e}_{1}$ and $\mathbf{e}_{2}$ are a pair of orthogonal
unit vectors in the coefficient space $\mathbb{R}^{N}$. Therefore,
the commutation relations for $N$ linear combination of position
and momentum can be characterized by a single real number, 
\begin{equation}
\mbox{Inc}(\mathbf{a},\mathbf{b})\equiv\left|\mathbf{a}\wedge\mathbf{b}\right|\,,\label{eq: degofInc}
\end{equation}
measuring the \emph{degree of incompatibility} of the observables
$\hat{r}_{j}$, $j=1\ldots N$. The relation \eqref{eq: CCR in standard form}
states that the original commutation relations are equivalent to a
situation in which all but two $\hat{r}$-observables have been mapped
to $0$, 
\begin{equation}
\hat{r}_{1}^{\prime}=\left|\mathbf{a}\wedge\mathbf{b}\right|{}^{\nicefrac{1}{2}}\,\hat{p}\,,\qquad\hat{r}_{2}^{\prime}=\left|\mathbf{a}\wedge\mathbf{b}\right|{}^{\nicefrac{1}{2}}\,\hat{q}\,,\qquad\hat{r}_{k}^{\prime}=0\,,\quad k=3\ldots N\,,\label{eq: r via a and b-1}
\end{equation}
corresponding to

\begin{equation}
\mathbf{a}^{\prime}=\left|\mathbf{a}\wedge\mathbf{b}\right|{}^{\nicefrac{1}{2}}\,\mathbf{e}_{1}\qquad\mbox{and\qquad}\mathbf{b}^{\prime}=\left|\mathbf{a}\wedge\mathbf{b}\right|{}^{\nicefrac{1}{2}}\,\mathbf{e}_{2}\,,\label{eq: standard a and b}
\end{equation}
respectively. We will obtain the standard form \eqref{eq: CCR in standard form}
by exploiting the fact that the norm of the bi-vector $\mathbf{a}\wedge\mathbf{b}$
is invariant under (i) gauge transformations and under (ii) transformations
of the vector operator $\hat{\mathbf{r}}$ which are orthogonal in
$\mathbb{R}^{N}$.

Before embarking on this calculation, we mention that other measures
of incompatibility for \emph{pairs} of observables exist. The\emph{
joint measurability region} \cite{busch+13b} quantifies the incompatibility
of two observables based on the amount of noise that needs to be added
in order for them to become jointly measurable. Based on this notion
a coarser measure can be introduced, the \emph{joint measurability
degree} \cite{heinosaari+14}, which returns a real number between
1/2 (corresponding to maximal incompatibility) and 1 (compatibility).
For continuous variables, the pair of position and momentum is found
to be maximally incompatible which agrees with the measure $\mbox{Inc}(\mathbf{a},\mathbf{b})$introduced
here. However, the case of three or more observables has not been
considered.

To derive the relation \eqref{eq: CCR in standard form}, we first
note that the observables
\begin{equation}
\hat{\mathbf{r}}_{U}=\hat{U}\,\hat{\mathbf{r}}\,\hat{U}^{\dagger}\,,\label{eq: general unitary trf of r}
\end{equation}
obtained from $\hat{\mathbf{r}}$ by any unitary operator $\hat{U}$,
satisfy the same commutation relations as the original observables
$\hat{\mathbf{r}}$. If we limit ourselves to \emph{linear }canonical
transformations of the observables $\hat{q}$ and $\hat{p}$, these transformations form the group $Sp(2,\mathbb{R})$, generated
by rotations, squeeze and gauge transformations described in \cite{arvind+95}.

Taking the unitary $\hat{U}=\hat{G}_{g}$ as defined in Eq. \eqref{eq: gauge operator},
position and momentum operators transform according to 
\begin{align}
\hat{p}_{g} & =\hat{p}\,,\nonumber \\
\hat{q}_{g} & =\hat{q}+g\hat{p}\,,\qquad g\in\mathbb{R}\,.\label{eq: p prime q prime gauge}
\end{align}
Clearly, the transformed coordinate vectors are $\mathbf{a}_{g}=\mathbf{a}$
and $\mathbf{b}_{g}=\mathbf{b}+g\mathbf{a}$. The components of the
vector operator $\hat{\mathbf{r}}_{g}$ have the same commutators
as those of $\hat{\mathbf{r}}$ as follows from the properties of
the exterior product, 
\begin{equation}
\mathbf{a}_{g}\wedge\mathbf{b}_{g}=\mathbf{a}\wedge\left(\mathbf{b}+g\mathbf{a}\right)=\mathbf{a}\wedge\mathbf{b}\,.\label{eq: rr =00003D00003D a wedge b gauge}
\end{equation}
Geometrically, the parameter $g$ labels a continuous family of parallelograms
with sides $\mathbf{a}_{g}$ and $\mathbf{b}_{g}$. They all have
the same area as they are related to each other by a shear transformation.
If the parameter $g$ takes the value 
\begin{equation}
g_{\perp}=-\frac{\mathbf{a}\cdot\mathbf{b}}{|\mathbf{a}|^{2}}\,,\label{eq: rectangle}
\end{equation}
the parallelogram turns into a \emph{rectangle} spanned by two \emph{orthogonal}
vectors, $\mathbf{a}_{\perp}=\mathbf{a}$ and $\mathbf{b}_{\perp}=\mathbf{b}+g_{\perp}\mathbf{a}$.

The right-hand-side of the commutation relations $\hat{\mathbf{r}}_{\perp}\wedge\hat{\mathbf{r}}_{\perp}=i\hbar\,\mathbf{a}_{\perp}\wedge\mathbf{b}_{\perp}\,\hat{I}$
now depends on the orthogonal vectors $\mathbf{a}_{\perp}$ and $\mathbf{b}_{\perp}$.
Denote unit vectors aligned with them by $\mathbf{e}_{a}$ and $\mathbf{e}_{b}$,
respectively, and consider an orthogonal transformation $\mathbf{R}$,
i.e. $\mathbf{R}\mathbf{R}^{T}=\mathbf{R}^{T}\mathbf{R}=\mathbf{I}$,
which rotates the vector operator $\hat{\mathbf{r}}_{\perp}$ into
\begin{equation}
\hat{\mathbf{r}}^{\prime}=\mathbf{R}\hat{\mathbf{r}}_{\perp}\,.\label{eq: map of r under R}
\end{equation}
Note that, typically, such a transformation \emph{cannot }be generated
by a unitary operator acting on the fundamental pair $\hat{p}$ and
$\hat{q}$. Since $\mathbf{e}_{a}\cdot\mathbf{e}_{b}=0$, we can always
find a transformation $\mathbf{R}$ which maps the vectors $\mathbf{e}_{a}$
and $\mathbf{e}_{b}$ to the first two elements of the standard basis,
\begin{equation}
\mathbf{e}_{a}=\mathbf{R}\mathbf{e}_{1}\,,\qquad\mathbf{e}_{b}=\mathbf{R}\mathbf{e}_{2}\,.\label{eq: map of e1 and e2 under R}
\end{equation}
The rotation $\mathbf{R}$ is unique only for $N=3$ since in $\mathbb{R}^{3}$
the map of the vectors $\mathbf{e}_{1}$ and $\mathbf{e}_{2}$ determines
the fate of the third basis vector, as $\mathbf{e}_{3}=\mathbf{e}_{1}\times\mathbf{e}_{2}$.
Using the definition of the outer product in \eqref{eq: def outer product}
and the fact that $\mathbf{R}^{-1}=\mathbf{R}^{T}$, one finds that
\begin{equation}
\left(\mathbf{R}\mathbf{a}\right)\otimes\left(\mathbf{R}\mathbf{b}\right)=\left(\mathbf{R}\mathbf{a}\right)\left(\mathbf{R}\mathbf{b}\right)^{T}=\mathbf{R}\left(\mathbf{a}\mathbf{b}^{T}\right)\mathbf{R}^{T}\,,\label{eq: def wedge-1}
\end{equation}
so that the exterior product \eqref{eq: def wedge} transforms according
to 
\begin{equation}
\left(\mathbf{R}\mathbf{a}\right)\wedge\left(\mathbf{R}\mathbf{b}\right)=\mathbf{R}\left(\mathbf{a}\mathbf{b}^{T}-\mathbf{b}\mathbf{a}^{T}\right)\mathbf{R}^{T}\equiv\mathbf{R}\left(\mathbf{a}\wedge\mathbf{b}\right)\mathbf{R}^{T}\,.\label{eq: def wedge-1-1}
\end{equation}
The relation \eqref{eq: Frobenius and bivector norm} now implies
that the length of the bi-vector $\mathbf{a}\wedge\mathbf{b}$ is
invariant under any rotation $\mathbf{R}$ applied to the $N$-component
vector operator $\hat{\mathbf{r}}$, 
\begin{equation}
\left|\left(\mathbf{R}\mathbf{a}\right)\wedge\left(\mathbf{R}\mathbf{b}\right)\right|^{2}=\frac{1}{2}\text{\mbox{Tr}}\left[\left(\mathbf{R}\mathbf{A}\mathbf{R}^{T}\right)^{T}\left(\mathbf{R}\mathbf{A}\mathbf{R}^{T}\right)\right]=\text{\ensuremath{\frac{1}{2}}\mbox{Tr}}\left(\mathbf{A}^{T}\,\mathbf{R}^{T}\mathbf{R}\,\mathbf{A}\,\mathbf{R}^{T}\mathbf{R}\right)=\left|\mathbf{a}\wedge\mathbf{b}\right|{}^{2}\,.\label{eq: frobenius for wedge-1}
\end{equation}
Applying this property to the vector operator $\hat{\mathbf{r}}^{\prime}=\mathbf{R}\left(\mathbf{a}_{\perp}\hat{p}+\mathbf{b}_{\perp}\hat{q}\right)$,
we finally obtain the desired result, Eq. \eqref{eq: CCR in standard form}.
In general, the vectors $\mathbf{a}_{\perp}$ and $\mathbf{\mathbf{b}_{\perp}}$
will be of different lengths. There is, however, a squeeze transformation
which rescales the pair $\hat{q},\hat{p}$, such that the lengths
of the vectors will be equal (see Appendix B).

\subsection{Maximal incompatibility}

The degree of incompatibility $\mbox{Inc}(\mathbf{a},\mathbf{b})$
defined in Eq. \eqref{eq: degofInc} can take any non-negative value.
If the vectors $\mathbf{a}$ and $\mathbf{b}$ are collinear, the
operators $\hat{r}_{j}$, $j=1\ldots N$, commute and hence are compatible,
$\mbox{Inc}(\mathbf{a},\lambda\mathbf{a})=0$, for all $\lambda\in\mathbb{R}$.
Multiplying the operators $\hat{r}_{j}$ by a common factor $\lambda\in\mathbb{R}$,
rescales their incompatibility accordingly, 
\begin{equation}
\mbox{Inc}(\lambda\mathbf{a},\lambda\mathbf{b})=\lambda\,\mbox{Inc}(\mathbf{a},\mathbf{b})\,.\label{eq: rescale Inc}
\end{equation}
To avoid artificially inflated values of incompatibility, it is natural
to require that the vectors $\mathbf{r}_{j}$ which fix the operators
$\hat{r}_{j}$, $j=1\ldots N$, have at most length one, 
\begin{equation}
|\mathbf{r}_{j}|^{2}\equiv r_{j}^{2}\leq1\,,\qquad j=1\ldots N\,.\label{eq: insidetheunitcircle}
\end{equation}
This constraint is consistent with Heisenberg's uncertainty relation
for the canonically conjugate pair of position and momentum observables.

What is the \emph{maximal }value which the\emph{
}incompatibility $\mbox{Inc}(\mathbf{a},\mathbf{b})=\left|\mathbf{a}\wedge\mathbf{b}\right|$
may take for $N$ observables $\mathbf{\hat{r}}$? The maximum is
of interest because it will determine the largest possible bounds
for the sum and the product inequalities, by ``exhausting'' the
quantum mechanical non-commutativity of the observables. Suppose we
are given $N$ observables defined by the vectors $\mathbf{r}_{j}=r_{j}\mathbf{u}_{j}$,
$j=1\ldots N$, where each $\mathbf{u}_{j}$ is a unit vector and
the lengths $r_{j}$ satisfy \eqref{eq: insidetheunitcircle}. Then,
the estimate 
\begin{equation}
\mbox{Inc}^{2}(\mathbf{a},\mathbf{b})=\sum_{j>k=1}^{N}A^{2}(\mathbf{r}_{j},\mathbf{r}_{k})=\sum_{j>k=1}^{N}r_{j}^{2}r_{k}^{2}A^{2}(\mathbf{u}_{j},\mathbf{u}_{k})\leq\sum_{j>k=1}^{N}A^{2}(\mathbf{u}_{j},\mathbf{u}_{k})\label{eq: Inc on boundary}
\end{equation}
shows that their incompatibility is smaller than that of $N$ observables
associated with the vectors $\mathbf{\widetilde{r}}_{j}=\mathbf{u}_{j}$,
with all their tips located on the unit circle. Thus, maximal incompatibility
will necessarily arise for an arrangement of $N$ points on the unit
circle.

It is instructive to discuss the simple case of $N=2$. Position $\hat{q}$
and momentum $\hat{p}$ satisfy Heisenberg's uncertainty relation
and should, of course, provide an example of maximal incompatibility.
The incompatibility of any two observables with vectors $\mathbf{r}_{j}=r_{j}\mathbf{u}_{j}$,$j=1,2$,
satisfying \eqref{eq: insidetheunitcircle} and with $\mathbf{u}_{1}\cdot\mathbf{u}_{2}=\cos\phi$,
is given by 
\begin{equation}
\mbox{Inc}^{2}(\mathbf{a},\mathbf{b})\equiv\left(a_{1}b_{2}-a_{2}b_{1}\right)^{2}=r_{1}^{2}r_{2}^{2}\sin^{2}\phi\leq1\,.\label{eq: Inc estimate}
\end{equation}
It achieves its maximum for $r_{1}=r_{2}=1$ and $\phi=\pm\pi/2$.
Thus, the pairs $\left(\hat{q},\pm\hat{p}\right)$ and all those obtained
from rotating them by an angle $\theta\in[0,2\pi$) indeed max out
the non-commutativity. The vectors $\mathbf{a}$ and $\mathbf{b}$
are necessarily orthogonal and of equal length. If the pair $\left(\mathbf{u}_{1},\mathbf{u}_{2}\right)$
describes a configuration with maximal incompatibility, then all four
configurations with vectors $\left(\pm\mathbf{u}_{2},\pm\mathbf{u}_{2}\right)$
are also maximally incompatible. We ignore the uncertainty preserving squeeze transformations here since they do not have
an equivalent for other values of $N$.

Let us now search for the arrangements of not just two but $N$ vectors
with tips on the unit circle which will result in maximal incompatibility.
Using the identity \eqref{eq: norm via a and b}, we find 
\begin{equation}
\mbox{Inc}(\mathbf{a},\mathbf{b})=|\mathbf{a}|\,|\mathbf{b}|\sin\phi\,,\qquad\phi\in[0,\pi)\,,\label{eq: Inc value from wedge product}
\end{equation}
where the angle between the two vectors in $\mathbb{R}^{N}$ is defined
by the relation $\mathbf{a}\cdot\mathbf{b}=|\mathbf{a}|\,|\mathbf{b}|\cos\phi$.
Summing the conditions $r_{j}^{2}=a_{j}^{2}+b_{j}^{2}=1$, $j=1\ldots N$,
over all values of $j$, one finds $|\mathbf{b}|^{2}=N-|\mathbf{a}|^{2}$
which implies 
\begin{equation}
\mbox{Inc}(\mathbf{a},\mathbf{b})=|\mathbf{a}|\,\sqrt{N-|\mathbf{a}|^{2}}\,\sin\phi\leq|\mathbf{a}|\,\sqrt{N-|\mathbf{a}|^{2}}\leq\frac{N}{2}\,.\label{eq: estimate for Inc}
\end{equation}
The last inequality follows because the function $f(x)=x\sqrt{N-x^{2}}$
has its unique maximum at $x=\sqrt{N/2}$. Thus, the incompatibility
takes the value $N/2$ if there exist $N$ observables characterized
by a pair $(\mathbf{a},\mathbf{b})$ of vectors which are orthogonal
and of equal length, $|\mathbf{a}|=|\mathbf{b}|=\sqrt{N/2}$.

According to Eq. \eqref{eq: polygon vectors a and b}, \emph{regular
polygons} with $N$ vertices located on the unit circle ($R_{N}\equiv1$)
correspond precisely to this situation. Thus, we may conclude that
the observables associated with the regular $N$-polygons introduced
in Sec.~\ref{NCanonical} maximize the incompatibility inherent in
$N$ observables linear in position and momentum. Clearly, this set
of observables is not the only one achieving the maximum: rotating the polygon by any angle in the interval $(0,2\pi/N)$ leads to
equivalent arrangements, as do individual reflections of the vectors
$\mathbf{r}_{j}$ about the origin.

\begin{figure}[t]
\centering \tikzset{>=latex} \begin{tikzpicture}[scale=1.4]\draw [ ->] (0,-2.7) -- (0,2.8);
\draw [ ->] (-2.7,0) -- (2.7,0);
\node at (0.3,2.7) {$\mathbf{p}$};
\node at (2.65,-0.3) {$\mathbf{q}$};
\node at (-0.2,2.2)  {$\mathbf{r}_1$};
\node at (-1.9,-1.2) {$\mathbf{r}_3$};
\node at (1.95,-1.2) {$\mathbf{r}_2$};
\draw [ ->] [line width=0.7mm] (0,0) -- (0,2);
\draw [ ->] [line width=0.7mm] (0,0) -- (1.73,-1);
\draw [ ->] [line width=0.7mm] (0,0) -- (-1.73,-1);
\draw[dashed] (0,0) circle (2cm);
\draw [dashed] [line width=0.5mm] (1.73,-1) -- (-1.73,1);
\draw [dashed] [line width=0.5mm] (-1.73,-1) -- (1.73,1);
\draw [dashed] [line width=0.5mm] (0,-2) -- (0,2);\end{tikzpicture} \caption{Phase-space visualization of three maximally incompatible observables:
each of the eight triples $(\pm\mathbf{r}_{1},\pm\mathbf{r}_{2},\pm\mathbf{r}_{3})$
corresponds to observables which maximise the incompatibility $\mbox{Inc}(\mathbf{a},\mathbf{b})$
since the variances $\Delta\hat{r}_{j}$ are invariant under $\hat{r}_{j}\to-\hat{r}_{j}$,
$j=1,2,3$. In addition, each configuration may be rotated rigidly
by any angle between $0$ and $2\pi/3$ without changing the value
of the incompatibility. For more than three observables, the equilateral
triangle with tips $(\mathbf{r}_{1},\mathbf{r}_{2},\mathbf{r}_{3})$
is replaced by a regular polygon with $N$ vertices. }
\label{fig: triples} 
\end{figure}
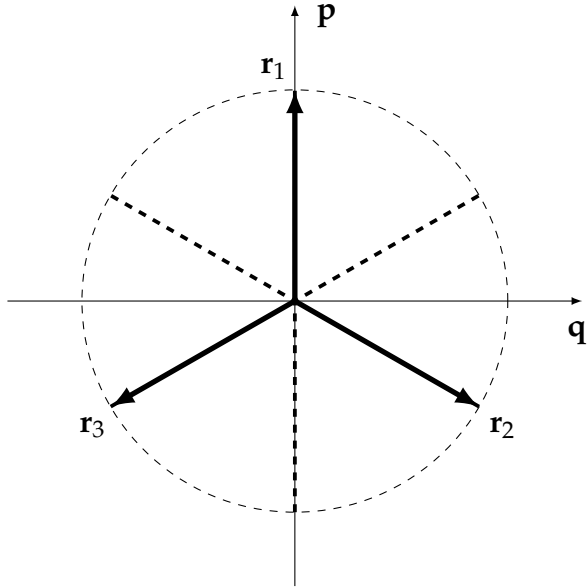

We suspect that no other sets of $N$ observables linear in position
and momentum will lead to maximal incompatibility. However, we are
only able to prove this property for $N=3$. Three observables as defined
in \eqref{eq: operator rj} associated with unit vectors $\mathbf{r}_{j}$
are conveniently parameterized by 
\begin{equation}
a_{j}=\cos\theta_{j}\,,\quad b_{j}=\sin\theta_{j}\,,\qquad\theta_{j}\in[0,2\pi)\,,\qquad j=1,2,3\,.\label{eq: coefficents for triple}
\end{equation}
Their incompatibility is given by a function of two variables, 
\begin{align}
\mbox{Inc}^{2}(\mathbf{a},\mathbf{b}) & =\sum_{j>k=1}^{3}(a_{j}b_{k}-a_{k}b_{j})^{2}=\sum_{j>k=1}^{3}\sin^{2}\left(\theta_{j}-\theta_{k}\right)\nonumber \\
 & =\frac{3}{2}-\frac{1}{2}\sum_{j>k=1}^{3}\cos\left(2\left(\theta_{j}-\theta_{k}\right)\right)\,.\label{eq: maximizing Inc}
\end{align}
Selecting the first observable to be momentum, $\hat{r}_{1}=\hat{p}$,
we have $\theta_{1}=0$. The maxima of the incompatibility occur when
one of the angles $\theta_{2}$ or $\theta_{3}$ takes the value $\pi/3$
or $4\pi/3$ while the other becomes $2\pi/3$ or $5\pi/3$. The solutions
for the observables $\hat{r}_{2}$ and $\hat{r}_{3}$ are shown in
Fig. \ref{fig: triples}, in terms of the vectors $\mathbf{r}_{j}$
characterizing them. It is straightforward to confirm that the vectors
$\mathbf{a}$ and $\mathbf{b}$ are indeed orthogonal for each set
of observables maximizing the incompatibility.

\section{Entropic uncertainty relations\label{Entropies}}

Heisenberg's uncertainty relation expresses a fundamental restriction
to simultaneously attribute specific values to both position and momentum
of a quantum particle. Hirschman \cite{hirschman57} used the position
and momentum probability densities of a quantum state $\ket{\psi}$
to capture this feature without referring to variances of observables.
Instead, he used the Shannon entropies\emph{ }of a state $\ket{\psi}$
associated with the modulus of the wave function in the position and
momentum representation. Given the state $\ket{\psi}$ with position
representation $\bk q{\psi}=\psi(q)$, its \emph{Shannon entropy}
\begin{equation}
S_{q}=-\int_{-\infty}^{\infty}dq\,|\psi(q)|^{2}\log\left(\sqrt{\hbar}|\psi(q)|^{2}\right)\,,\label{eq: Shannon position}
\end{equation}
returns small values for probability densities $|\psi(q)|^{2}$ which
are localized and large ones for densities which are spread out; the
factor $\sqrt{\hbar}$ ensures that the argument of the logarithm
is dimensionless. The momentum representation of the state $\ket{\psi}$
follows from Fourier-transforming its position wave function, 
\begin{equation}
\bk p{\psi}=\psi(p)=\frac{1}{\sqrt{2\pi\hbar}}\int_{-\infty}^{\infty}e^{-ipq/\hbar}\psi(q)\,dq\,,\label{eq: momentum rep wave function}
\end{equation}
leading to a probability density $|\psi(p)|^{2}$ with Shannon entropy
$S_{p}$, defined in analogy to Eq. \eqref{eq: Shannon position}.
Hirschman showed that the sum of these entropies cannot fall below
zero and conjectured that a tighter nonzero bound would hold,\footnote{This form of Hirschman's inequality holds if one sets a free dimensionless
parameter equal to one as explained in Ref. \cite{barchielli+17}.} 
\begin{equation}
S_{q}+S_{p}\geq\ln(e\pi)\,.\label{entropicUR-1}
\end{equation}
Using the properties of a norm for the Fourier transform \cite{babenko61,beckner75},
this uncertainty relation has been proved in \cite{bialnicki+75},
nearly 20 years after being conjectured.

The inequalities by Hirschman and Heisenberg are related closely.
The variance of the observable $\hat{p}_{\theta}=\hat{p}\cos\phi+\hat{q}\sin\phi$,
$\phi\in[0,2\pi)$, has a lower bound \cite{shannon+49} 
\begin{equation}
\Delta^{2}p_{\phi}\geq\frac{\hbar}{2e\pi}e^{2S_{\phi}}\,,\label{eq: variance entropy estimate}
\end{equation}
which depends on the Shannon entropy associated with the probability
density $|\bk{p_{\phi}}{\psi}|^{2}=|\psi(p_{\phi})|^{2}$, where $\hat{p}_{\phi}\ket{p_{\phi}}=p_{\phi}\ket{p_{\phi}}$
holds. Using Eq. \eqref{eq: variance entropy estimate} for both momentum
and position (i.e. for $\phi=0$ and $\phi=\pi/2$, respectively),
the entropic inequality \eqref{entropicUR-1} indeed implies 
\begin{equation}
\Delta^{2}p\,\Delta^{2}q\,\geq\left(\frac{\hbar}{2e\pi}\right)^{2}e^{2\left(S_{0}+S_{\pi/2}\right)}\geq\left(\frac{\hbar}{2}\right)^{2}\,,\label{eq: variance-entropic Heisenberg}
\end{equation}
as already pointed out by Hirschman \cite{hirschman57}. If the system
resides in the ground state of a harmonic oscillator with unit mass
and frequency, i.e. in the coherent state $\ket 0$, we have $\Delta^{2}p=\Delta^{2}q=\hbar/2$.
Both inequalities in \eqref{eq: variance-entropic Heisenberg} are
now saturated since Eq. \eqref{eq: variance entropy estimate} turns
into an equality (which happens whenever the state is represented
by a Gaussian \cite{bialnicki+75}) so that $S_{p}=S_{q}=(1/2)\ln\left(e\pi\right)$.
In other words, the value of the tight bound in Hirschman's inequality
\eqref{entropicUR-1} is obtained if one considers a case in which
the pair product-uncertainty relation is saturated and combines it
with the bound \eqref{eq: variance entropy estimate}.

This argument does not, of course, replace the proof of Hirschman's
inequality. However, we use an analogous argument to conjecture a
bound for a generalization of Hirschman's inequality which involves
more than two observables linear in position and momentum. Consider
$N\geq3$ observables $\hat{r}_{j}=\hat{p}\cos\phi_{j}+\hat{q}\sin\phi_{j}$,
$\phi_{j}=2\pi(j-1)/N$, $j=1\ldots N$, associated with a regular
$N$-polygon with vertices on the unit circle. The product inequality
\eqref{eq: simple SumN and Prod N} is known to be saturated if the
system resides in the state $\ket 0$, 
\begin{equation}
\prod_{j=1}^{N}\Delta^{2}r_{j}=\left(\frac{\hbar}{2}\right)^{N}\,.\label{eq: ProdN equality}
\end{equation}
As the wave function of the state $\ket 0$ is Gaussian in each $\hat{r}_{j}$-representation,
we have 
\begin{equation}
\Delta^{2}r_{j}=\frac{\hbar}{2e\pi}e^{2S_{j}}\,,\qquad j=1\ldots N\,,\label{eq: variance entropy identity for rj}
\end{equation}
where $S_{j}$ is the Shannon entropy of the probability density $|\psi(r_{j})|^{2}$
of the state $\ket 0$. Substituting \eqref{eq: variance entropy identity for rj}
into \eqref{eq: ProdN equality}, we find that 
\begin{equation}
\frac{2}{N}\sum_{j=1}^{N}S_{j}=\ln\left(e\pi\right)\label{eq: Gaussian equality for N obs}
\end{equation}
holds, leading to the conjecture of an $N$-observable Hirschman-type
inequality, 
\begin{equation}
\frac{2}{N}\left(S_{1}+S_{2}+\ldots+S_{N}\right)\geq\ln\left(e\pi\right)\,,\quad N\geq3.\label{entropicN}
\end{equation}
Other inequalities exist for the case of the $r$-observables defined
by vertices distributed \emph{inhomogeneously }on the unit circle
since these configurations result in a smaller degree of incompatibility.

\section{Summary and discussion}

In this paper we have derived inequalities for $N$ linear combinations
of position and momentum of a quantum particle. The sum and the product
inequality, Eqs. \eqref{eq: SumNG} and \eqref{ProdNG}, depend on
one single parameter only, the \emph{degree of incompatibility} $\mbox{Inc}(\mathbf{a},\mathbf{b})$
defined in Eq. \eqref{eq: degofInc}. This number is the only relevant
parameter once the original $N(N-1)$ commutator relations \eqref{eq: rr =00003Da wedge b}
have been brought to the standard form \eqref{eq: CCR in standard form}.

Using the relation between the arithmetic and the geometric mean,
we can concatenate the two inequalities, 
\begin{equation}
\frac{\Delta^{2}r_{1}+\Delta^{2}r_{2}+\ldots+\Delta^{2}r_{N}}{N}\geq\left(\Delta^{2}r_{1}\,\Delta^{2}r_{2}\cdots\Delta^{2}r_{N}\right)^{\nicefrac{1}{N}}\geq\frac{\hbar\left|\mathbf{a}\wedge\mathbf{b}\right|}{N}\,,\label{eq: ArithGeoCombinationOf ProdSum}
\end{equation}
neatly summarizing our main findings for the variances of multiple
observables linear in position and momentum, valid for arbitrary (pure
or mixed) quantum states. Given the product inequality, the bound
of the geometric mean by the arithmetic mean actually provides an
alternative derivation of the sum inequality. Heisenberg's inequality
and the triple inequality emerge as the first two members of a family
labeled by $N=2,3,\ldots$ The cases $N=2$ and $N=3$ are special
since they are the only ones in which \emph{all} pairwise commutators
can be made to coincide.

Upon rescaling the observables by a common positive factor, $\hat{r}_{j}\to\hat{r}_{j}\sqrt{\left|\mathbf{a}\wedge\mathbf{b}\right|}$,
$j=1\ldots N$, the inequalities \eqref{eq: ArithGeoCombinationOf ProdSum}
take a particularly simple form, 
\begin{equation}
\frac{\Delta^{2}r_{1}+\Delta^{2}r_{2}+\ldots+\Delta^{2}r_{N}}{N}\geq\left(\Delta^{2}r_{1}\,\Delta^{2}r_{2}\cdots\Delta^{2}r_{N}\right)^{\nicefrac{1}{N}}\geq\frac{\hbar}{N}\,,\label{eq: ArithGeoCombinationOf ProdSum-1}
\end{equation}
showing immediately that saturation occurs if each variance takes
the value $\hbar/N.$ For a square, i.e. the regular polygon with $N=4$ vertices, there are additional states which saturate the second inequality but not the first one: the existence of this one-parameter family of squeezed states is a direct consequence of the well-known invariance of Heisenberg's uncertainty relation (i.e. $N=2$) under squeezing transformations. 

To identify $N$ linear observables $\hat{r}_{j}$ with \emph{maximal}
incompatibility, we have considered sets characterized by vectors
$\mathbf{r}_{j}$,$j=1\ldots N$, of unit length or less. In this
case, the bound on the right-hand-side of Eq. \eqref{eq: ArithGeoCombinationOf ProdSum}
reaches its maximum whenever the $N$-dimensional coefficient
vectors satisfy the condition $\left|\mathbf{a}\wedge\mathbf{b}\right|=N/2$.
This happens, for example, if the vectors $\pm\mathbf{r}_{j}$,$j=1\ldots N$,
are of unit length and their tips form a regular polygon in $\mathbb{R}^{2}$
(for a suitable choice of signs). The bound \eqref{eq: ArithGeoCombinationOf ProdSum}
takes the value \emph{zero} if the coefficient vectors satisfy $\mathbf{a}=\lambda\mathbf{b}$,
where $\lambda\in\mathbb{R}$. Consequently, all $N$ observables
will be scalar multiples of each other and hence commute, corresponding
to arrangements of \emph{minimal} incompatibility.

Furthermore, we conjectured entropic inequalities to hold for more
than two continuous variables, analogous in form to the relation originally
discovered by Hirschman. The sum of the Shannon entropies associated
with $N$ directions in phase space is expected to achieve its maximum
if the angles between any neighboring directions equal $2\pi/N$.
We expect that there will be no states violating the conjectured bound
\eqref{entropicN} which has been derived from evaluating the $N$
Shannon entropies in a Gaussian state. This $N$-term generalization
of Hirschman's inequality fills a gap concerning entropic inequalities
for \emph{continuous }variables while in \emph{finite-dimensional}
Hilbert spaces numerous investigations of entropic inequalities for
multiple variables have been carried out already.

Our results raise a number of questions which we hope to address in
future work. Let us begin by pointing out a surprising formal similarity
between the result \eqref{eq: ArithGeoCombinationOf ProdSum} and
the inequality for the sum of standard deviations of two spin observables
\cite{busch+13}: 
\begin{equation}
\Delta A+\Delta B\geq|\mathbf{A}\times\mathbf{B}|\,,\label{eq: sum of two spin variances}
\end{equation}
where $\hat{A}=\mathbf{A}\cdot\hat{\boldsymbol{\sigma}}$ and $\hat{B}=\mathbf{B}\cdot\hat{\boldsymbol{\sigma}}$,
with unit vectors $\mathbf{A},\mathbf{B}\in\mathbb{R}^{3}$, and $\hat{\boldsymbol{\sigma}}=\left(\hat{\sigma}_{x},\hat{\sigma}_{y},\hat{\sigma}_{z}\right)^{T}$
is a vector operator with Pauli matrices as components. Here, the
vectors $\mathbf{A}$ and $\mathbf{B}$ collect coefficients of \emph{different}
observables, hence should be compared to the vectors $\mathbf{r}_{j}$,
$j=1\ldots N$, and not to the coefficient vectors $\mathbf{a}$ and
$\mathbf{b}$, respectively. Is there a simple generalization of \eqref{eq: sum of two spin variances}
valid for the sum of the standard deviations of more than two spin
observables? Since the observables $\hat{A},\hat{B},\ldots$ will
be in a one-to\--one-\-corres\-pon\-dence with $N$ points inside
of the unit sphere, a natural bound on the incompatibility of $N$
observables is likely to define a geometric structure in $\mathbb{R}^{3}$,
just as regular polygons in $\mathbb{R}^{2}$ emerge in the case of
$N$ continuous variables.

To conclude, we discuss our results from a fundamental perspective.
Heisenberg's uncertainty relation has often been understood to say
that one cannot simultaneously associate definite values to both position
and momentum of a quantum particle. Kochen-Specker-type arguments
\cite{kochen+69} formalize this insight by showing that non-contextual
value-assignments are algebraically \textendash{} i.e. not statistically
\textendash{} at odds with quantum predictions. Contradictions arise
from dichotomic observables for both discrete \cite{mermin+90,peres91}
and continuous quantum variables \cite{clifton00,myrvold02}. A recent
probabilistic approach \cite{klyachko+08} introduces non-contextual
``Kochen-Specker inequalities'' which lend themselves to experimental
verification. Our results may have implications for similar contextuality
arguments given in terms of phase-space translations, along the lines
of Refs. \cite{plastino+10,asadian+15}, for example.

\subsection*{Acknowledgements}

S. K. acknowledges financial support by the Greek \emph{State Scholarship
Foundation }(IKY) as well as the \emph{WW Smith Fund,} held by the
Departments of Mathematics and Physics at the University of York,
where part of this work was conducted. S. W. appreciates helpful discussions with and suggestions by Paul
Busch and Roger Colbeck about the Shannon entropy for continuous variables.
Finally, the authors would like to thank an unknown referee for pointing
out the direct derivation of the inequality \eqref{eq:linear UR}
for mixed states.

\appendix

\section*{Appendix A\label{sec:Appendix-A}}

In this appendix, we derive that the sum inequality \eqref{eq: SumNG}
also holds for mixed states, the validity of \ref{eq:linear UR}
for pure states being our point of departure. To do so, we first consider
the sum of the variances of observables $\hat{A}_{1},\hat{A}_{2},\ldots$,
and derive the inequality
\begin{align}
\sum_{j}\Delta_{\rho}^{2}A_{j}\geq\sum_{j}\Delta_{\psi}^{2}A_{j}\,,\label{eq: mixed>pure}
\end{align}
where $\hat{\rho}$ is any mixed state and $\ket{\psi}$ is a pure
state which will depend on $\hat{\rho}$. This relation implies that
it is sufficient to consider pure states only when searching for universal
bounds on sums of variances.

Let $\hat{A}$ be a self-adjoint operator and suppose that the mixed
state $\hat{\rho}=\lambda\hat{\rho}_{1}+(1-\lambda)\hat{\rho}_{2}$
is a convex combination of two density matrices $\hat{\rho}_{1}$
and $\hat{\rho}_{2}$, with $\lambda\in[0,1]$. Then, the variance
of $\hat{A}$ in the mixture $\hat{\rho}$ is bounded from below by
the sum of the variances of the in the states $\hat{\rho}_{1}$ and
$\hat{\rho}_{2}$, i.e. 
\begin{align}
\Delta_{\rho}^{2}A\geq\lambda\Delta_{\rho_{1}}^{2}A+(1-\lambda)\Delta_{\rho_{2}}^{2}A\,,\label{concavity of variance}
\end{align}
as follows from the concavity of the variance.

To prove that the variance $\Delta_{\rho}^{2}A=\avg{A^{2}}_{\rho}-\avg A_{\rho}^{2}$,
with $\langle\hat{A}\rangle_{\rho}\equiv\text{Tr}\left(\hat{A}\hat{\rho}\right)$
etc., \emph{is} concave, we note that 
\begin{align}
\avg{\hat{A}^{2}}_{\rho}=\lambda\avg{\hat{A}^{2}}_{\rho_{1}}+(1-\lambda)\avg{\hat{A}^{2}}_{\rho_{2}}\,,\label{first term}
\end{align}
and 
\begin{align}
\avg{\hat{A}}_{\rho}^{2}=\left(\lambda\avg{\hat{A}}_{\rho_{1}}+(1-\lambda)\avg{\hat{A}}_{\rho_{2}}\right)^{2}\leq\lambda\avg{\hat{A}}_{\rho_{1}}^{2}+(1-\lambda)\avg{\hat{A}}_{\rho_{2}}^{2}\,,\label{second term}
\end{align}
using the convexity of the function $f(x)=x^{2}$. The inequalities
\eqref{first term} and \eqref{second term} immediately imply inequality
\eqref{concavity of variance}. Since one of the variances on the
right-hand-side of \eqref{concavity of variance}, say $\Delta_{\rho_{1}}^{2}A$,
must be less than or equal to the left-hand-side, we obtain
\begin{align}
\Delta_{\rho}^{2}A\geq\Delta_{\rho_{1}}^{2}A\,.
\end{align}

This argument can be extended to a sum of the variances of $N$ Hermitean
operators $\hat{A}_{j},j=1\ldots N$, resulting in
\begin{align}
\sum_{j}\Delta_{\rho}^{2}A_{j}\geq\sum_{j}\Delta_{\rho_{1}}^{2}A_{j}\,.\label{concavity for more than two}
\end{align}
To see this, note that the sum of two concave functions (such as
$\Delta_{\rho}^{2}A$ and $\Delta_{\rho}^{2}B$) is also concave which
leads to 
\begin{align}
\Delta_{\rho}^{2}A+\Delta_{\rho}^{2}B & \geq\lambda\left(\Delta_{\rho_{1}}^{2}A+\Delta_{\rho_{1}}^{2}B\right)+(1-\lambda)\left(\Delta_{\rho_{2}}^{2}A+\Delta_{\rho_{2}}^{2}B\right)\,,\label{concavity of sum}
\end{align}
for any mixture $\hat{\rho}=\lambda\hat{\rho}_{1}+(1-\lambda)\hat{\rho}_{2}$.
Again, one of the two terms in brackets on the right-hand-side of
\eqref{concavity of sum} is less than or at most equal to the left-hand-side.
Thus, we have shown that the inequality \eqref{concavity for more than two}
holds for two operators. It is straightforward to include more operators. 

Finally, to complete the proof of \eqref{eq: mixed>pure}, we need
to consider a state with a convex decomposition given by to $\hat{\rho}=\sum_{k}r_{k}\hat{P}_{k}$,
where the operators $\hat{P}_{k}=\kb{\psi_{k}}{\psi_{k}}$, $k=1,2,\ldots$,
project onto pure states $\ket{\psi_{k}}$. Then, for the variance
of a single observable $\hat{A}$ we have the bound 
\begin{align}
\Delta_{\rho}^{2}A\geq\sum_{k}r_{k}\Delta_{\psi_{k}}^{2}A\geq\Delta_{\psi}^{2}A\,,
\end{align}
where $\ket{\psi}$ is one of the states $\ket{\psi_{1}},\ket{\psi_{2}},\ldots$,
for which the variances of the right-hand-side falls below or is equal
to the left-hand-side. Since this argument also applies to the sum
of variances of observables $\hat{A}_{1},\hat{A}_{2},\ldots$, the
inequality \eqref{eq: mixed>pure} does indeed hold. 

\appendix

\section*{Appendix B\label{sec: Appendix-B}}

In this Appendix, we will show that the product $\mathbf{a}\wedge\mathbf{b}$
is invariant (i) under any phase-space rotation of the position and
momentum observables and (ii) under squeezing transformations.

(i) Consider any rotation of position $\hat{q}$ and momentum $\hat{p}$
in phase space, 
\begin{align}
\hat{p}_{\vartheta} & =\hat{p}\cos\vartheta+\hat{q}\sin\vartheta\,,\nonumber \\
\hat{q}_{\vartheta} & =-\hat{p}\sin\vartheta+\hat{q}\cos\vartheta\,,\qquad\vartheta\in[0,2\pi]\,.\label{eq: p prime q prime}
\end{align}
This commutator-preserving transformation is generated by the unitary
\begin{equation}
\hat{R}_{\vartheta}=\exp\left[-i\vartheta\left(\hat{p}^{2}+\hat{q}^{2}\right)/2\hbar\right]\,,\label{eq: unitary generating rotations}
\end{equation}
known as the time-evolution operator of a harmonic oscillator with
unit mass and frequency. The relations \eqref{eq: p prime q prime}
induce linear transformations in the coefficient space $\mathbb{R}^{N}$,
which you obtain upon replacing the symbols $\hat{p}$ and $\hat{q}$
in Eq. \eqref{eq: p prime q prime} by $\mathbf{a}$ and $\mathbf{b}$,
respectively. Therefore, the exterior product of the transformed vectors
reads 
\begin{equation}
\mathbf{a}_{\vartheta}\wedge\mathbf{b}_{\vartheta}=\left(\mathbf{a}\cos\vartheta+\mathbf{b}\sin\vartheta\right)\wedge\left(-\mathbf{a}\sin\vartheta+\mathbf{b}\cos\vartheta\right)=\mathbf{a}\wedge\mathbf{b}\,,\label{eq: rr =00003D00003D a wedge b-1}
\end{equation}
confirming the expected invariance.

(ii) Rescaling the observables $\hat{p}$ and $\hat{q}$ is achieved
by the unitary operator $\hat{U}=\hat{S}_{\gamma}$ (see Eq. \eqref{Squeeze})
which \emph{squeezes} the momentum and position operators according
to 
\begin{align}
\hat{p}_{\gamma} & =\gamma\hat{p}\,,\nonumber \\
\hat{q}_{\gamma} & =\frac{1}{\gamma}\hat{q}\,,\qquad\gamma\neq0\,.\label{eq: p prime q prime scale}
\end{align}
It is easy to see that the coefficient vectors in $\mathbb{R}^{N}$
transform in a covariant way, namely 
\begin{align}
\mathbf{a}{}_{\gamma} & =\gamma\mathbf{a}\,,\nonumber \\
\mathbf{b}_{\gamma} & =\frac{1}{\gamma}\mathbf{b}\,,\qquad\gamma\neq0\,,\label{eq: p prime q prime scale-1}
\end{align}
which implies the invariance of the product, $\mathbf{a}_{\gamma}\wedge\mathbf{b}_{\gamma}=\mathbf{a}\wedge\mathbf{b}$.
Choosing the value 
\begin{equation}
\gamma_{0}=\left(\frac{|\mathbf{b}|}{|\mathbf{a}|}\right)^{\nicefrac{1}{2}},\label{eq: value for gamma}
\end{equation}
allows us to introduce new coefficient vectors $\mathbf{a}$ and $\mathbf{b}$
with equal length given by$\left(|\mathbf{a}|\,|\mathbf{b}|\right)^{\nicefrac{1}{2}}\equiv\left|\mathbf{a}\wedge\mathbf{b}\right|{}^{\nicefrac{1}{2}}$. 
\end{document}